\documentclass[conference]{IEEEtran}
\IEEEoverridecommandlockouts
\usepackage{cite}
\usepackage{amsmath,amssymb,amsfonts}
\usepackage{algorithmic}
\usepackage{graphicx}
\usepackage{textcomp}
\usepackage{xcolor}
\usepackage{subcaption}
\usepackage{color,soul}
\usepackage{booktabs}
\usepackage{multirow}
\usepackage{comment}

\usepackage[hyphens]{url}

\def\BibTeX{{\rm B\kern-.05em{\sc i\kern-.025em b}\kern-.08em
    T\kern-.1667em\lower.7ex\hbox{E}\kern-.125emX}}
\begin{document}

\title{\Huge Benchmarking Optimizers for Qumode State Preparation with Variational Quantum Algorithms}

\author{

Shuwen Kan\textsuperscript{\rm 1*},
Miguel Palma\textsuperscript{\rm 1*}\thanks{*Authors contributed equally},
Zefan Du \textsuperscript{\rm 1}\\
Samuel A Stein\textsuperscript{\rm 2},
Chenxu Liu\textsuperscript{\rm 2}, 
Juntao Chen\textsuperscript{\rm 1},
Ang Li\textsuperscript{\rm 2},
and Ying Mao\textsuperscript{\rm 1}
\\

 \textsuperscript{\rm 1} Computer and Information Science Department, Fordham University, \\ \{sk107, mip2, zdu19, jchen504, ymao41\}@fordham.edu\\
    \textsuperscript{\rm 2}Pacific Northwest National Laboratory (PNNL), \{samuel.stein, chenxu.liu, ang.li\}@pnnl.gov\\
}

\maketitle

\begin{abstract}
Quantum state preparation involves preparing a target state from an initial system, a process integral to applications such as quantum machine learning and solving systems of linear equations. Recently, there has been a growing interest in qumodes due to advancements in the field and their potential applications. However there is a notable gap in the literature specifically addressing this area. This paper aims to bridge this gap by providing performance benchmarks of various optimizers used in state preparation with Variational Quantum Algorithms. We conducted extensive testing across multiple scenarios, including different target states, both ideal and sampling simulations, and varying numbers of basis gate layers. Our evaluations offer insights into the complexity of learning each type of target state and demonstrate that some optimizers perform better than others in this context. Notably, the Powell optimizer was found to be exceptionally robust against sampling errors, making it a preferred choice in scenarios prone to such inaccuracies. Additionally, the Simultaneous Perturbation Stochastic Approximation optimizer was distinguished for its efficiency and ability to handle increased parameter dimensionality effectively. 

\end{abstract}

\section{introduction}

Quantum computing has the potential to solve computational problems that are simply intractable on classical computers\cite{shor1998quantum}. Superposition, an exponentially large computational Hilbert space, and entanglement, give rise to a new model of computation. Seminal algorithms such as Shor's Algorithm \cite{shor1999polynomial} or Digital Quantum Simulation of physical systems\cite{abrams1997simulation}, motivate the field to develop quantum computers powerful enough to tackle these problems.

Despite this significant potential, hardware and software problems limit the physical realization of quantum computers\cite{preskill2018quantum}. Arguably, the largest challenge in quantum computing is the inability to escape noise\cite{resch2021benchmarking}. Imperfect control, decoherence, and undesirable coupling, are problems that plague quantum computing platforms today but quantum error correction provides a path towards exponential error suppression \cite{roffe2019quantum}.



Quantum computing diverges from today's classical computers, which rely solely on transistor-based technology, as researchers have not yet standardized a single approach for constructing the perfect qubit. Currently, the field features multiple architectures, each with its own set of strengths and weaknesses. Superconducting circuits, for instance, utilize supercooled circuits to create an anharmonic oscillator \cite{Upadhyay2022ArchitecturesFQ}, offering rapid operation times. However, these systems face challenges such as short decoherence times and limited qubit connectivity. On the other hand, trapped ion quantum computers employ lasers to manipulate and control charged atoms held in place by magnetic fields \cite{bruzewicz2019trapped}. While these systems boast longer decoherence times and are generally more manageable, they operate more slowly and present difficulties in optical and electrical control.



Qumodes represent a promising approach in quantum computing, characterized by their use of continuous variable systems over infinite dimensional matrices \cite{Braunstein_2005}. Typically associated with bosonic modes, these systems theoretically can reach infinitely high Fock levels. However, such levels are not physically realistic, prompting the truncation of these matrices to a manageable, finite dimension that reflects the highest achievable Fock level. Despite being less explored than other quantum systems, qumodes are particularly noteworthy due to their natural alignment with real-world applications. Examples include optimizing graph cliques \cite{banchi2020molecular}, analyzing point processes \cite{PhysRevE.101.022134}, and simulating molecular vibrational spectra \cite{huh2015boson}, showcasing their potential in practical scenarios.
There have been notable developments in qumode control and readout~\cite{Lu_2023, zhou2024bosehedral, fabre2020modes, oberdieck2023quantum}. Beamsplitter operations with gate fidelities exceeding 99.98\%\cite{Lu_2023} have been reported and show promise that high-reliability physical implementations are possible in the near future. Additionally, a compiler optimzation framework for bosonic quantum acheived a 39.6\% reduction of Beamsplitter gate counts while maintaining fidelity at 98.00\% \cite{zhou2024bosehedral}, improving the qumode software and compiler state of the art.

Variational Quantum Algorithms (VQA) have garnered interest due to their ability to mitigate the effects of noise in contemporary quantum hardware \cite{stein2022quclassi, Cerezo_2021, stein2021qugan, stein2022eqc, stein2022qucnn, zhao2020measurement, mu2022iterative, baheri2021tqea, self2021variational, d2023distributed, l2024quantum}. VQAs operate by utilizing quantum hardware to execute parameterized quantum circuits, alongside classical systems that run optimization algorithms to adjust these parameters. This dual-system approach enables the application of VQAs in a variety of fields, including solving optimization and eigenvalue problems that are prevalent in quantum physics and chemistry simulations.



State preparation is a crucial subroutine in various quantum algorithms, notably in Variational Quantum Algorithms (VQA)-based quantum learning, Hamiltonian simulation, and solving systems of linear equations \cite{biamonte2017quantum}. This process, which involves loading classical information into quantum devices before computation, is notably resource-intensive. Specifically, preparing N-dimensional vectors necessitates a circuit with depth proportional to O(N) \cite{aaronson2015read}. For practical applications to realize Quantum Advantage, the development of an efficient state preparation method is paramount. While traditional quantum state preparation has been extensively studied \cite{smith2021efficient, zhang2022quantum, lin2020near, sun2023asymptotically}, qumode state preparation has received less attention.

In this paper, we focus on VQA-based applications and investigate the quantum learning of a gate sequence tailored for arbitrary qumode target states, aimed at enhancing qumode state preparation efficiency. The key contributions of this research are as follows:
\begin{itemize}

    \item We conduct a comprehensive benchmark study on the performance of various optimizers in qumode state preparation with different settings.

    \item Our evaluations compare various optimizers in the task of Local and Non-local Gaussian states as well as Non Gaussian using a 2-gateset VQA.
    
    \item The experiments demonstrate that learning Non-Gaussian states posed a greater challenge, requiring up to 10 layers to minimize state preparation error.
\end{itemize}

\section{Related Work}

When discussing how quantum computers outperform classical devices, a crucial component is the loading of classical data onto these quantum devices. The No Cloning Theorem states that that arbitrary quantum states cannot be duplicated the same way data can be duplicated on classical devices \cite{wootters2009no}, making it a non-trivial problem. The key problem can be stated as determining the correct quantum gate sequence that transforms an initial quantum state to the desired target state with minimal error and in the least amount of time possible.

Quantum state preparation schemes are classified into two broad categories: arithmetic decomposition (AD) and variational quantum state preparation (VQSP) \cite{wang2023robuststate}. AD approaches employ a more analytical approach to generating circuits to produce the target state, however they are affected by quantum noise on real devices that throw off the results. VQSP, on the other hand, uses the variational method to approximate the target state by iteratively updating a set of parameters. This approach is more suited to NISQ devices because of its limited circuit depth and gate count.

Variational quantum algorithms (VQA) utilize classical optimizers to determine which direction to adjust parameters. Optimizers are also organized under two categories: gradient-based and gradient-free optimizers. The gradient is the derivative of the objective function that specifies the rate of change of input to the output. Gradient-free methods sample the function at various points to derive an estimated value of the gradient. The choice of optimizer to use affects the quality of the output and the number of iterations needed to approach a reasonable approximate solution.

Bosonic quantum is a separate branch of quantum computing based on bosonic qumodes. The fundamental difference is that these qumodes are continuous variable systems, as opposed to the 2-state qubit system. However, this capacity for an infinite number of states is truncated when simulating on classical devices. Bosonic quantum has immense potential to advance quantum algorithms because it presents a simpler and more versatile way of encoding and representing quantum information \cite{RevModPhys.84.621}.

Despite the potential of bosonic qumodes and the necessity of state preparation in quantum algorithms, there is barely any research on preparation of target qumode states. A study shows that an energy-dependent barren plateau phenomenon is exhibited in the cost function of training bosonic VQCs for state preparation \cite{zhang2023energydependent}. Our contribution is to provide an empirical benchmark and comparison of different optimizers for this task.


\section{background}

\subsection{Qumodes}

Qumodes are a type of quantum system characterized by having an infinite number of levels, which makes them capable of being in a superposition of multiple Fock basis states. These states are represented by a linear combination: $\rho = \alpha_0|0\rangle + ... + \alpha_n|N\rangle$, where $\alpha_0$, ... , $\alpha_n$ are complex numbers that sum to 1 when squared. When dealing with systems comprising multiple qumodes, their overall state can be described by the tensor product of the individual qumodes' states. However, in practical applications, qumodes are often truncated to a manageable number of Fock levels, ranging from 0 (the vacuum state) to $N$, where $N$ serves as the cutoff value. This truncation simplifies calculations and makes experimental realizations feasible.

\subsection{Variational Qumode Circuits}
Variational Quantum Algorithms (VQAs) are hybrid algorithms that employ both quantum and classical computing techniques to find approximate solutions to problems. These algorithms are particularly effective in mitigating the limitations of current quantum devices, which include limited qubit counts and high susceptibility to noise. In VQAs, a parametrized quantum circuit encodes the problem at hand, while a classical optimization algorithm iteratively adjusts the circuit's input parameters to refine the solution. However, it is important to note that VQAs do not guarantee the discovery of an exact solution.

The effectiveness of VQAs in practical applications is often hindered by several limitations. One significant challenge is the Barren Plateau (BP) phenomenon, where the landscape of the cost function is extremely flat across vast regions, with the solution residing in a narrow, difficult-to-locate gorge \cite{Cerezo_2021}. In such scenarios, the optimizer struggles to find adequate gradient information to effectively navigate out of the BP, often resulting in prolonged convergence times or the acceptance of suboptimal solutions. The expressiveness of the parametrized circuit and the choice of initial parameters are crucial factors that influence the likelihood of encountering BPs.


\subsection{Optimizers}



Optimizers play a crucial role in Variational Quantum Algorithms (VQAs) as they navigate the cost landscape to minimize the cost function. Integral to logistics planning, operations research, and machine learning, these algorithms are categorized into two primary groups: gradient-based methods and gradient-free methods. This classification highlights the varied approaches used to tackle optimization problems in quantum computing environments.

The performance of these optimizers is assessed on several fronts. Accuracy is crucial as it measures how closely the optimizer's solution approaches the global minimum and its robustness against noise, essential for reliable quantum computing. Speed, determined by the number of iterations to convergence and the number of samples per iteration, reflects the efficiency of the optimizer. Scalability indicates the optimizer’s performance as the number of parameters increases, important for complex systems. Lastly, consistency evaluates the optimizer's ability to deliver similar results across different runs with randomized initial parameters, ensuring reliability and repeatability in outcomes. Together, these factors determine the effectiveness of optimizers in practical quantum computing applications.

\section{evaluation}

This section evaluates the performance of various optimizers for VQAs on qumode systems, specifically in learning parameter values to approximate arbitrary target states. The parameters are structured into layers, with each layer comprising five real numbers: $[v_r,v_i,\theta_x, \theta_y, \theta_z]$. Here, $v_r + v_i*i$ forms the complex number for the VP gate, and $\theta_x$, $\theta_y$, $\theta_z$ serve as input parameters for the RX, RY, and RZ gates, respectively.

For the experiments, we implement testing cases with Qiskit Bosonic~\cite{stavenger2022c2qa} along with customized gates and QuTiP functions. This study employs the Python libraries Scipy~\cite{scipySciPy} and SPSA~\cite{pennylaneOptimizationUsing}, renowned for their extensive use in scientific computing and engineering. We have tested following optimizers including L-BFGS-B~\cite{wikipediaLimitedmemoryBFGS}, Conjugate Gradient (CG)~\cite{wikipediaConjugateGradient}, Sequential Least Squares Programming (SLSQP)~\cite{readthedocshostedSLSQPx2014}, Constrained Optimization by Linear Approximation (COBYLA)~\cite{scipyScipyoptimizefmin_cobylax2014}, Nelder-Mead~\cite{ibmNELDER_MEADQuantum}, Powell~\cite{scipyMinimizemethodx2019Powellx2019x2014}, Implicit Filtering (ImFil)~\cite{ibmIMFILQuantum}, Bound Optimization BY Quadratic Approximation (PyBobyQA)~\cite{ibmBOBYQAQuantum}, and Simultaneous Perturbation Stochastic Approximation (SPSA)~\cite{wikipediaSimultaneousPerturbation}.

Additionally, to evaluate the similarity between the learned and target states, we utilize the SWAP test. Given the prohibitive cost of state tomography, the SWAP test provides a more efficient means of approximating state similarities due to its lower sampling requirements. In our study, which focuses on learning latent states and reproducing them parametrically, the SWAP test is particularly suitable. It measures the squared inner product between two qumode state vectors, translating this measure into the probability of observing a '0' on an ancilla coupled transmon, as determined by the algorithm.

\subsection{System Setup}
\subsubsection{Basis gates}
Given a qubit-coupled bosonic quantum system, we opt to analyse applying layers of following computationally complete sets of instructions:
\[ V_p(\alpha) = e^{i\sigma_z(\alpha a^\dagger - \alpha^* a)} \]
\[ R(\hat{b},\theta) = e^{i\theta \hat{b} \cdot \hat{\sigma}} \]
\begin{itemize}
    \item \( \alpha \) is a complex number for $V_p$ gate
    \item \( \theta \) is a real number in the range \( [0,2\pi) \) for rotation,
    \item \( \hat{b} \) is a unit 3D vector,
    \item \( \hat{\sigma} = (\sigma_x, \sigma_y, \sigma_z) \) represents the Pauli matrices.
\end{itemize}

Layers are comprised of a complete U3 rotation on the coupled transmon parameterised by two parameters, providing complete control over the transmon, followed by an application of the parametric displation gate (Vp). This is considered a layer in further discussion.

\subsubsection{Objective Function}
Following the gate layer applications, the swap test is performed between the two qumodes. The swap test results in the transmon having a P($|0\rangle$) of $0.5 + 0.5 \times |\langle\Phi\Psi\rangle|^2$. Our goal is to maximize the fidelity of our learnt state and a target state, hence optimizing directly over the swap test results can lead to gradient scaling issues due to the squared fidelity term. To address this, we applied a square root transformation to the swap test results, resulting in $1 - \sqrt{2 \times (P(|0\rangle) - 0.5)}$. Consequently, our final objective function becomes infidelity. In scenarios where sampling simulation produces swap test results lower than 0.5, leading to negative values inside the square root, a lower bound was imposed to ensure that the swap test result remains greater than 0.5.

\subsubsection{Target State of Interest}
Given the impracticality of simulating infinite dimensions, we truncate the bosonic systems to a 10-level system for general experiments, which we deemed sufficient to represent the distributions of interest. Three types of target states were explored in our experiments: Local Gaussian , Gaussian, and Non-Gaussian states. We selected an identical target state for different optimizers under the same settings. The target states we investigated were as follows: Local Gaussian (mean = 0, std = 0.75), Gaussian distribution (mean = 5, std = 1), and Non-Gaussian distribution (0, 0.209, 0.417, 0.209, 0, 0.417, 0.626, 0.417, 0, 0).
\subsubsection{Workload}
In this study, two primary metrics, the final infidelity value and the number of function evaluations (nfev) are evaluated for benchmarking purpose. Infidelity directly relates to optimizer's ability to approach global optimal. The number of function evaluations directly relates to the runtime efficiency for each optimizer, correlating to how many times one would need to optimize over real life experiments. In the result table, a standard deviation of 0 may be observed for COBYLA and SPSA. This occurs because Scipy enforces a 1000-iteration limit for the COBYLA optimizer, causing certain results to display an nfev standard deviation of 0. Furthermore, a custom iteration count of 1000 is utilized for the SPSA optimizer. We conducted experiments with varying numbers of layers and truncation levels. Each experimental setup was replicated 30 times.  

Each target state is tested by two simulation methods, ideal simulation and sampling simulation. In ideal simulation, the swap test results reflects the use of the statevector directly, whereas sampling simulation incorporates a sampling process, representing the real world experimental process.

\subsubsection{Optimizers}
The optimizers we tested fall into two categories: gradient-based optimizers and derivative optimizers. Moreover, to effectively tackle the challenges arising from the high-dimensional nature of the problem when multiple layers are needed to approximate the target state closely, we utilized the Simultaneous Perturbation Stochastic Approximation (SPSA) optimizer.

In the context of gradient-based optimizers, the selection of step size for finite difference gradient estimation is crucial. Smaller step sizes tend to be heavily influenced by sampling error, while larger sample sizes may be able to provide an unbiased estimator of the gradient. Throughout our experiments, we discovered that an adaptive step size approach was prone to encountering local optima. Consequently, we opted for a fixed step size of 0.03 for ideal simulation. However, for sampling simulation, both the step size and sampling size must be carefully addressed to ensure accurate results. Table \ref{tab:step_sample_evaluation} illustrates the percentage of non-converging trials for L-BFGS-B, a gradient-based optimizer, across various sampling sizes and step sizes. We conducted 200 trials with sample sizes ranging from 1024 to 8192, with increments of 1024 and step size selection of [0.03,0.05,0.08].

The findings indicate that a sampling size of 1024 proves insufficient for providing a meaningful gradient estimation direction, resulting in non-converging trials ranging from 15\% to 38.5\%. As sampling size reaches between 6144 and 8192, percentage of non-converging dropped below 5\%. Given that increasing the sampling size corresponds to an increase in the number of shots of measurement in practical implementation, which can be costly, we opted to use a sampling size of 6144 for our subsequent sampling simulations. 

It is notable that in ideal simulation, as indicated in Table \ref{tab:LG_NS}, the L-BFGS-B optimizer is expected to attain an infidelity of 0.003. For trials that converge, the final infidelity is anticipated to be close to this value, albeit subject to some sampling error. Hence, the standard deviation is closely associated with the number of non-converging trials, potentially yielding values within the range of [0.3, 0.9].

\begin{table}
  \centering
    \caption{Evaluation of Various Step Sizes and Sample Sizes for gradient-based optimizer for Sampling Simulation}
    \label{tab:step_sample_evaluation}
   \scalebox{0.8}{
    \begin{tabular}{rrrrrr}
    \toprule
    trial & sample\_size & step size & infidelity\_mean & infidelity\_std & None\_converge(\%) \\

    \midrule
     200 &1024 & 0.03 & 0.299 & 0.408 & 38.5\\
     200 &1024 & 0.05 & 0.198 & 0.354 & 25\\
     200 &1024 & 0.08 & 0.131 & 0.296 & 15\\
     
     200 &2048 & 0.03 & 0.210 & 0.362 & 25.5\\
     200 &2048 & 0.05 & 0.126 & 0.292 & 14.5\\
     200 &2048 & 0.08 & 0.113 & 0.281 & 11.5\\

     200 &3072 & 0.03 & 0.199 & 0.360 & 23\\
     200 &3072 & 0.05 & 0.0950 & 0.252 & 12\\
     200 &3072 & 0.08 & 0.0629 & 0.194 & 7\\

     200 &4096 & 0.03 & 0.198 & 0.361 & 23\\
     200 &4096 & 0.05 & 0.0604 & 0.184 & 9.5\\
     200 &4096 & 0.08 & 0.0418 & 0.115 & 5.5\\

     200 &5120 & 0.03 & 0.0844 & 0.233 & 10\\
     200 &5120 & 0.05 & 0.0510 & 0.159 & 5.5\\
     200 &5120 & 0.08 & 0.0427 & 0.128 & 7\\

     200 &6144 & 0.03 & 0.143 & 0.317 & 17\\
     200 &6144 & 0.05 & 0.0593 & 0.192 & 6\\
     1000 &6144 & 0.08 & 0.0332 & 0.192 & 4.8\\
     1000 &7168 & 0.08 & 0.0290 & 0.0948 & 3.5\\
     1000 & 8192 & 0.08 & 0.0297 & 0.106 & 3.5\\
    \bottomrule
    \end{tabular}}
\end{table}
 
\subsection{Local Gaussian State}
\subsubsection{Ideal Simulation}
The results can be found in Table \ref{tab:LG_NS} and our we observe that the Powell optimizer was able to minimize infidelity the most at the cost of also having the highest average number of function evaluations. The mean and std dev of infidelity remains more or less consistent even when the cutoff and layer count values are changed while the mean and std dev of the number of evaluations showed a lot more variation in the results. There is no observed relationship between the number of layers and cutoff on the average infidelity.

\begin{table}[ht]
    \centering
    \caption{Local Gaussian State of Ideal Simulation}
    \label{tab:LG_NS}
        \scalebox{0.8}{
        \begin{tabular}{rrrrrr}
        \toprule
        layers & method & infidelity\_mean & infidelity\_std & nfev\_mean & nfev\_mean \\
        \midrule
        1  & CG & 0.00358 & 0.00192 & 34.58 & 22.08\\
        2  & CG & 0.00309 & 0.000722 & 140.4 & 65.32\\
        \hline
        1  & L-BFGS-B &  0.00353 & 0.00308 & 21.34 & 15.17\\
        2  & L-BFGS-B &  0.00385 & 0.00504 & 53.66 & 25.93\\
        \hline
        1  & SLSQP &  0.00405 & 0.00208 & 18.24 & 14.31\\
        2  & SLSQP &  0.00355 & 0.000948 & 23.48 & 10.86 \\
        \hline
        1  & SPSA &  0.00740 & 0.0111& \multirow{2}{*}{1000} & \multirow{2}{*}{0}\\
        2 & SPSA &  XXXX& XXXX \\
        \hline
        1  & Nelder-Mead &  0.00313 & 0.000442 & 254.9 & 60.77\\
        2  & Nelder-Mead &  0.00300 & 0.000609 & 991.28 & 366.13\\
        \hline
        1  & Powell &  0.00330 & 0.000713 & 487.82 & 282.64\\
        2  & Powell &  0.00339 & 0.000567 & 2415.06 & 1617.42\\
        \hline
        1  & COBYLA &  0.00369 & 0.00156 & 427.3 & 452.93\\
        2 & COBYLA &  0.00425 & 0.000920 & 952.58 & 189.93\\
        \bottomrule
        \end{tabular}}
\end{table}

\subsubsection{Sampling Simulation}

The results are displayed in Table \ref{tab:LG_S}. The Powell optimizer showcases the highest robustness against sampling errors. Generally, derivative-free optimizers outshine gradient-based ones, as sampling error markedly impacts the gradient estimation process. Among the three gradient-based optimizers, CG is the most susceptible to sampling error, while SPSA exhibits the most robustness. Notably, increasing the layer from 1 to 2 for the CG optimizer helps achieve results comparable to other gradient-based optimizers.
\begin{table}
    \centering
    \caption{Local Gaussian State of Sampling Simulation}
    \label{tab:LG_S}
    \scalebox{0.8}{
        \begin{tabular}{rrrrrr}
        \toprule
        layers & method & infidelity\_mean & infidelity\_std & nfev\_mean & nfev\_mean  \\
        \midrule
        1 &CG&  0.156 & 0.162 & 32.85 & 8.526\\
        2 &CG&  0.0188 & 0.0168 & 35.8 & 10.272\\
        \hline
        1  &  L-BFGS-B &  0.0337  & 0.0336 & 17.2 & 5.205 \\
        2  &  L-BFGS-B &  0.0142 & 0.0122 & 16.1 & 4.433  \\
        \hline
        1  & SLSQP &  0.111 & 0.115 & 21.05 & 8.565 \\
        2  & SLSQP &  0.0152 & 0.0116 & 27 & 14.259 \\
        \hline
        1  & SPSA &  0.0112 & 0.0161 & \multirow{2}{*}{1000} & \multirow{2}{*}{0}\\
        2 & SPSA &  0.00957 & 0.0119 \\
        \hline
        1  & Nelder-Mead & 0.0943 & 0.0977 & 2773.15 & 67.729 \\
        2  & Nelder-Mead &  0.0797 & 0.209 & 2854.6 & 95.060 \\
        \hline
        1  & Powell &  0.00516 & 0.00450 & 183.5 & 64.209 \\
        2  & Powell & 0.00778 & 0.00608 & 491.9 & 147.625 \\
        \hline
        1  & COBYLA &  0.0154 & 0.0143 & 61.9 & 4.959 \\
        2  & COBYLA &  0.0250 & 0.0203 & 108.3 & 9.934 \\
    \bottomrule
    \end{tabular}}
\end{table}

\subsection{Gaussian State}
\subsubsection{Ideal Simulation}
The results are displayed in Table \ref{tab:G_NS}.
With an increased number of layers, the dimensionality of parameters expands, creating a more intricate optimization space. Consequently, the number of function evaluations grows nonlinearly with each additional layer. An unusual trend observed in this table is that for layer=1, CG and L-BFGS-B optimizers require a significantly higher number of function evaluations. This anomaly arises because, with only one layer, the final learned state fails to capture the full complexity of the Gaussian distribution. Consequently, these two optimizers struggle to converge. Prior to termination, they expend additional iterations attempting to achieve convergence.

It's important to note that Powell consistently outperforms other optimizers despite this increase in complexity. While derivative-free optimizers require a greater number of function evaluations, they maintain efficiency in terms of time. It's crucial to acknowledge that gradient-based optimizers (CG, L-BFGS-B, SLSQP) entail additional objective function evaluations for finite difference gradient estimation. This process involves evaluating the objective function multiple times based on the dimensionality of the parameter space. For example, with two layers, each gradient estimation requires 20 function evaluations. Therefore, the overall number of objective function evaluations for CG, L-BFGS-B, and SLSQP is 2499, 882, and 672, respectively. With three layers, these numbers increase to 7920, 2759, and 1984, respectively, with CG being the most inefficient. Since the number of objective function evaluations directly affects the optimizer's runtime, gradient-based optimizers become less efficient as the number of layers increases, making derivative-free optimizers more favorable in terms of time efficiency.

\begin{table}[ht]
    \centering
    \caption{Gaussian State of Ideal Simulation}
    \label{tab:G_NS}
    \scalebox{0.8}{
        \begin{tabular}{rrrrrr}
        \toprule
        layers & method & infidelity\_mean & infidelity\_std & nfev\_mean & nfev\_mean \\
        \midrule
        1 &CG&  0.372 & 0.198 & 642.8 & 288.16\\
        2 &CG&  0.128 & 0.0516 & 119.56 & 28.47\\
        3 &CG&  0.0549 & 0.0489 & 264.7 & 94.92\\
        \hline
        1  & L-BFGS-B  & 0.361 & 0.194 & 375.38 & 187.62\\
        2  &  L-BFGS-B &  0.140 & 0.0599 & 42.7 & 16.68\\
        3  &  L-BFGS-B &  0.0564 & 0.0501 & 88.9 & 37.05\\
        \hline
        1  & SLSQP &  0.363 & 0.196 & 115.8 & 33.50\\
        2  & SLSQP &  0.131& 0.0516 & 32.1& 9.7\\
        3  & SLSQP &  0.0502& 0.0492 & 63.75& 31.31\\
        \hline
        1  & SPSA &  0.384 & 0.227 & \multirow{3}{*}{1000} & \multirow{3}{*}{0}\\
        2& SPSA & 0.180 & 0.0517 \\
        3 & SPSA & 0.119 & 0.0681 \\
        \hline
        1  & Nelder-Mead &  0.382 & 0.200 & 227.05 & 41.73\\
        2  & Nelder-Mead &  0.121 & 0.0438 & 1124.36 & 229.32\\
        3  & Nelder-Mead &  0.0922 & 0.0406 & 1324.7 & 17.61\\
        \hline
        1  & Powell &  0.250 & 0.106 & 413.4 & 150.75\\
        2  & Powell &  0.136 & 0.0552 & 2193.3 & 1101.89\\
        3  & Powell &  0.0468 & 0.0274 & 8540.75 & 3671.44\\
        \hline
        1  & COBYLA &  0.313 & 0.173 & 80.7 & 10.56\\
        2  & COBYLA &  0.142 & 0.0575 & 994.84 & 36.12\\
        3  & COBYLA &  0.0977 & 0.0351 & 1000 & 0\\ 
        \bottomrule
        \end{tabular}}
\end{table}

\begin{figure*}[htbp]
  \centering
  \begin{subfigure}[b]{0.23\textwidth}
    \includegraphics[width=\textwidth]{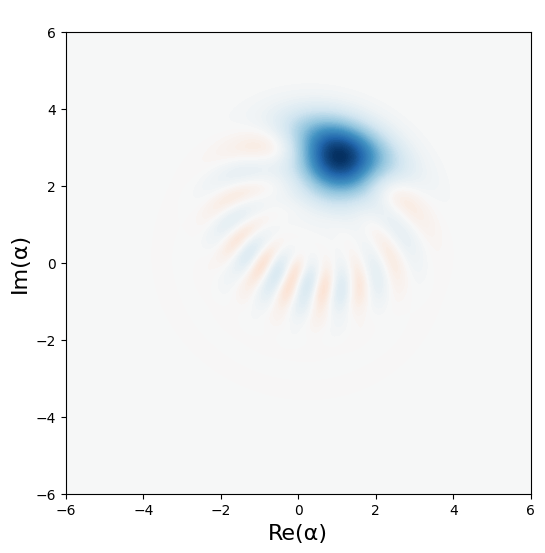}
    \caption{Learnt State layer = 1}
    \label{fig:sub1}
  \end{subfigure}
  \begin{subfigure}[b]{0.23\textwidth}
    \includegraphics[width=\textwidth]{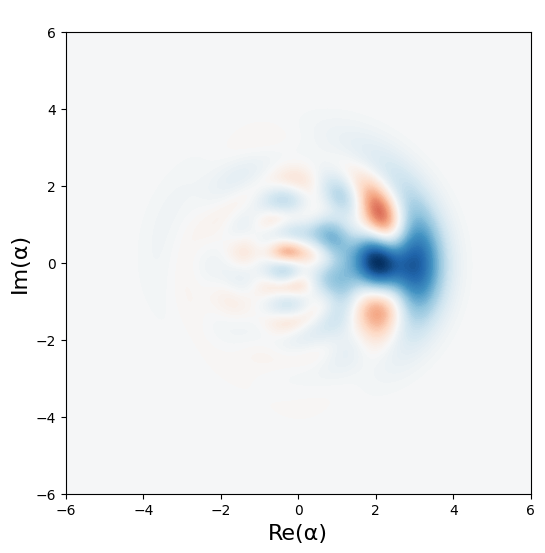}
    \caption{Learnt State layer = 5}
    \label{fig:sub2}
  \end{subfigure}
  \begin{subfigure}[b]{0.23\textwidth}
    \includegraphics[width=\textwidth]{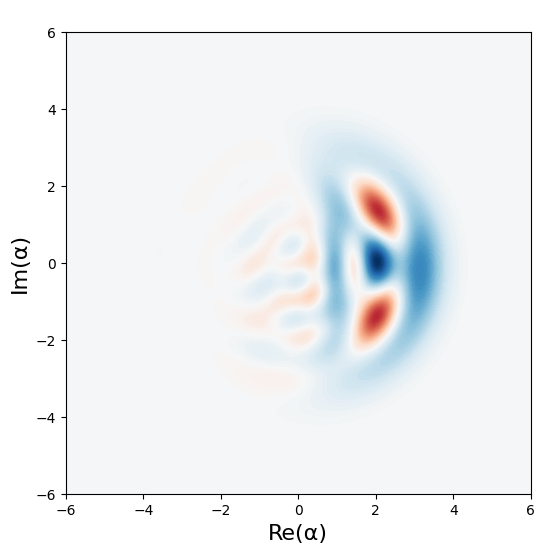}
    \caption{Learnt State layer = 10}
    \label{fig:sub3}
  \end{subfigure}
  \begin{subfigure}[b]{0.23\textwidth}
    \includegraphics[width=\textwidth]{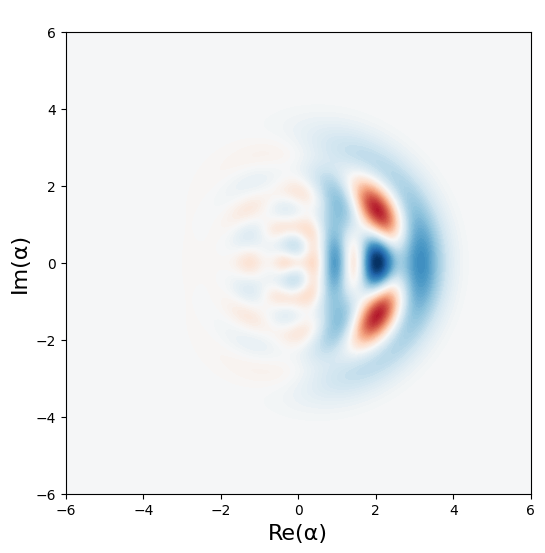}
    \caption{target state}
    \label{fig:sub4}
  \end{subfigure}
  \caption{Wigner Function of Learnt Stat at different Layer for Non-Gaussian State}
  \label{fig:main}
\end{figure*}

\subsubsection{Sampling Simulation}
The results are displayed in Table \ref{tab:G_S}. It's noteworthy that Nelder-Mead displays the lowest resilience to sampling errors, consistent with findings for both local Gaussian and non-Gaussian states. Powell continues to demonstrate superior performance among all optimizers, with SPSA also showing favorable results. Despite this, all optimizers yield final infidelity values of only 0.11 and 0.13 for 3 layers. Hence, achieving convergence in sampling simulation for learning Gaussian states requires more than 3 layers.
\begin{table}
    \centering
    \caption{Gaussian State of Sampling Simulation}
    \label{tab:G_S}
    \scalebox{0.8}{
        \begin{tabular}{rrrrrr}
        \toprule
        layers & method & infidelity\_mean & infidelity\_std & nfev\_mean & nfev\_mean  \\
        \midrule
        1 &CG&  0.518 & 0.274 & 32.15 & 9.65\\
        2 &CG&  0.198 & 0.0638 & 40.95 & 14.03\\
        3 &CG&  0.179 & 0.0693 & 41 & 13.05\\
        \hline
        1  &  L-BFGS-B & 0.395 & 0.179 & 18.2 & 9.16 \\
        2  &  L-BFGS-B &  0.234 & 0.0742 & 17.55 & 4.87\\
        3 &L-BFGS-B&  0.170 & 0.0877 & 26.13 & 9.14\\
        \hline
        1  & SLSQP &  0.414 & 0.198 & 42.05 & 19.15 \\
        2  & SLSQP &  0.208 & 0.0677 & 67.25 & 32.64\\
        3 &SLSQP&  0.176 & 0.0549 & 37.47 & 13.52\\
        \hline
        1  & SPSA &  0.341 & 0.183 & \multirow{3}{*}{1000} & \multirow{3}{*}{0}\\
        2& SPSA & 0.175 & 0.0533 \\
        3 & SPSA & 0.118 & 0.0511 \\
        \hline
        1  & Nelder-Mead &  0.659 & 0.299 & 2849.95 & 408.62 \\
        2  & Nelder-Mead &  0.429 & 0.223 & 2816.55 & 71.52\\
        3 &Nelder-Mead&  0.447 & 0.259 & 2955.13 & 157.06\\
        \hline
        1  & Powell &  0.242 & 0.0179 & 189.8 & 68.24 \\
        2  & Powell &  0.183 & 0.0531 & 538.25 & 245.94\\
        3 &Powell&  0.134 & 0.0503 & 1153.27 & 505.55\\
        \hline
        1  & COBYLA &  0.372 & 0.0978 & 64.2 & 9.71 \\
        2  & COBYLA &  0.225 & 0.0896 & 117.7 & 10.921\\
        3 &COBYLA&  0.215 & 0.0894& 163.8 & 11.37\\
    \bottomrule
    \end{tabular}}
\end{table}

\subsection{Non-Gaussian State}
\subsubsection{Ideal Simulation}
Learning a non-Gaussian state presents inherent challenges compared to learning Gaussian states due to its more complex structure. Gaussian states typically have simpler parameterizations based on mean and standard deviation, while non-Gaussian states may necessitate more intricate parameterizations.

Another challenge associated with increasing layers is the resultant increase in the number of parameters. Testing beyond 3 layers, particularly for gradient-based optimizers, becomes computationally expensive. However, SPSA stands out as an exception, reducing the number of function evaluations required for each gradient estimation to just two, regardless of dimensionality. Consequently, to investigate the required number of layers for learning non-Gaussian states, we tested layers 1 to 10 using the SPSA optimizer. The results shown in Table \ref{tab:NG_NS} indicate a nearly linear relationship for the SPSA optimizer from 1 layer to 10 layers.

\subsubsection{Sampling Simulation}
\begin{figure}
    \centering
    \includegraphics[width=0.45\textwidth]{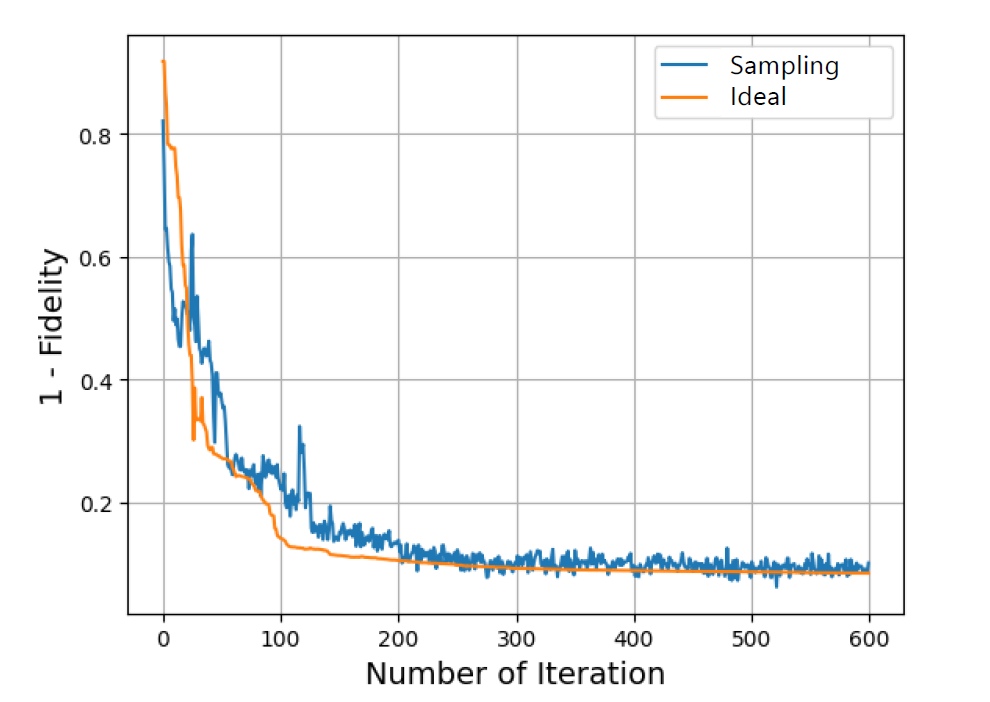} 
    \caption{Stochastic and non-Stochastic simulation}
    \label{fig:example}
\end{figure}

Figure \ref{fig:example} presents a comparison of the infidelity change over iterations for SPSA in both ideal and sampling simulations for layer = 10. It highlights the characteristics of sampling simulation, where the objective function exhibits slight fluctuations due to sampling errors.

Similar to the ideal simulation, the error decreases almost linearly and stabilizes as the number of layers approaches 10. In summary, our findings suggest that employing 10 layers is adequate to capture the features of the non-Gaussian state.

Furthermore, Figure \ref{fig:main} presents the Wigner function visualization to illustrate the similarity between the learned and target states across various layers. It distinctly depicts that with 1 layer, the learned state retains the shape of a local Gaussian distribution. At 5 layers, corresponding to an infidelity value of 0.125, the basic shape of a non-Gaussian state begins to emerge. Finally, with 10 layers and an infidelity value of 0.04, the learned state closely resembles the target state. These plots also aid in understanding how the infidelity value correlates with the Wigner representation of Qumodes.
 
\begin{table}
    \centering
    \caption{Non-Gaussian State of Ideal Simulation}
    \label{tab:NG_NS}
    \scalebox{0.8}{
        \begin{tabular}{rrrrrr}
        \toprule
        layers & method & infidelity\_mean & infidelity\_std & nfev\_mean & nfev\_mean  \\
        \midrule
        1 &CG&  0.340 & 0.187 & 656.45 & 274.83 \\
        2 &CG&  0.223 & 0.0964 & 2803.6& 758.29\\
        3 &CG&  0.112 & 0.0373 & 8209 & 2055.25\\
        \hline
        1  & L-BFGS-B  &  0.340 & 0.187 & 335.5 & 187.34\\
        2  &  L-BFGS-B &  0.221 & 0.108 & 1173.9 & 489.98 \\
        3 & L-BFGS-B&  0.142 & 0.0353 & 3633.2 & 1456.66\\
        \hline
        1  & SLSQP &  0.341 & 0.189 & 113.45& 28.14\\
        2  & SLSQP &  0.205 & 0.0756 & 539.4 & 201.61 \\
        3 &SLSQP&  0.158 & 0.0305 & 1658.6 & 835.49\\
        \hline
        1  & Nelder-Mead &  0.361 & 0.196 & 240.45 & 43.37\\
        2  & Nelder-Mead &  0.218 & 0.129 & 1025.85 & 306.71 \\
        3 &Nelder-Mead&  0.158 & 0.0252 & 1328.5 & 19.60\\
        \hline
        1  & Powell &  0.240 & 0.0892 & 319.85 & 124.35\\
        2  & Powell &  0.170 & 0.0200 & 1769 & 774.26 \\
        3  & Powell &  0.148 & 0.0323 & 5774.4 & 2825.30\\
        \hline
        1  & COBYLA & 0.302 & 0.167 & 82.35 & 12.24\\
        2  & COBYLA & 0.205 & 0.113 & 941.8 & 106.44 \\
        3 &COBYLA&  0.194 & 0.0542 & 1000 & 0\\
        \hline
        1 & SPSA & 0.480 & 0.0396 & \multirow{10}{*}{1000} & \multirow{10}{*}{0}\\
        2 & SPSA & 0.348 & 0.00937\\
        3 & SPSA & 0.290 & 0.0111 \\
        4 & SPSA & 0.238 & 0.00330\\
        5 & SPSA & 0.125 & 0.00330 \\
        6 & SPSA & 0.104 & 0.000678\\
        7 & SPSA & 0.0780 & 0.00102 \\
        8 & SPSA & 0.0494 & 0.000446\\
        9 & SPSA & 0.0408 & 0.000294 \\
        10 & SPSA & 0.0413 & 0.000179\\
    \bottomrule
    \end{tabular}}
\end{table}

\begin{table}
    \centering
    \caption{Non-Gaussian State of Sampling Simulation}
    \scalebox{0.8}{
        \begin{tabular}{rrrrrr}
        \toprule
        layers & method & infidelity\_mean & infidelity\_std & nfev\_mean & nfev\_mean  \\
        \midrule
        1 &CG&  0.672 & 0.202 & 38.8 & 14.30\\
        2 &CG&  0.555 & 0.205 & 33.1 & 14.21\\
        3 &CG&  0.310 & 0.150 & 41.8 & 20.26\\
        \hline
        1  &  L-BFGS-B &  0.618 & 0.225 & 24.05 & 8.83 \\
        2  &  L-BFGS-B &  0.552 & 0.209 & 23 & 11.23\\
        3  &  L-BFGS-B &  0.356 & 0.185 & 19.4 & 3.95\\
        \hline
        1  & SLSQP &  0.661 & 0.226 & 32.9 & 12.85 \\
        2  & SLSQP &  0.503 & 0.200 & 62.05 & 29.61 \\
        3  & SLSQP &  0.337 & 0.207 & 32.4 & 13.43\\
        \hline
        1  & Nelder-Mead & 0.574 & 0.267 & 2757.65 & 59.85 \\
        2  & Nelder-Mead &  0.468 & 0.232 & 2871.3 & 91.10 \\
        3  & Nelder-Mead &  0.412 & 0.196 & 2931.5 & 120.39\\
        \hline
        1  & Powell &  0.277 & 0.146 & 258.1 & 111.74 \\
        2  & Powell &  0.198 & 0.0263 & 563.05 & 260.35 \\
        3  & Powell &  0.208 & 0.0449 & 820 & 424.66\\
        \hline
        1  & COBYLA &  0.402 & 0.200 & 64 & 12.43 \\
        2  & COBYLA &  0.294 & 0.136 & 107.5 & 11.11 \\
        3  & COBYLA &  0.251 & 0.0761 & 157.2 & 20.00\\
        \hline
        1 & SPSA & 0.474 & 0.0147 & \multirow{10}{*}{1000} & \multirow{10}{*}{0}\\
        2 & SPSA & 0.417 & 0.0167\\
        3 & SPSA & 0.270 & 0.00319\\
        4 & SPSA & 0.218 & 0.00137\\
        5 & SPSA & 0.158 & 0.00603 \\
        6 & SPSA & 0.111 & 0.000829\\
        7 & SPSA & 0.0839 & 0.000943 \\
        8 & SPSA & 0.0567 & 0.000399\\
        9 & SPSA & 0.0546 & 0.000165 \\
        10 & SPSA & 0.0455 & 0.000497\\
    \bottomrule
    \end{tabular}}
\end{table}

\section{conclusion}


This paper evaluates various optimizers for preparing target qumode states, focusing on three types: Local Gaussian, Non-local Gaussian, and Non-Gaussian states. Each state type is evaluated with specific parameters, including step size and sampling size. In ideal simulations, both Local and Non-local Gaussian states were effectively approximated within three layers. However, simulations involving sampling required additional layers to achieve similar accuracy. The preparation of Non-Gaussian states proved more challenging, necessitating up to 10 layers to minimize the state preparation error effectively. The study also delves into the implications of increasing the number of layers on runtime efficiency and optimization success. It reveals that additional layers lead to more iterations needed to find optimal points within an expanded parameter space. Furthermore, the analysis shows that gradient-based optimizers tend to struggle more with increases in dimensionality compared to derivative-free optimizers. An exception is the SPSA optimizer, which maintains efficient gradient estimation capabilities irrespective of the system's dimensionality. This insight is crucial for optimizing quantum state preparation processes, particularly in more complex quantum systems.

\section{Acknowledgement}
 This research was supported in part by the National Science Foundation (NSF) under grant agreements 2329020, 2301884 and 2335788. This work was partially supported by the U.S. Department of Energy, Office of Science, National Quantum Information Science Research Centers, Co-design Center for Quantum Advantage (C2QA) under contract number DE-SC0012704, (Basic Energy Sciences, PNNL FWP 76274). This research used resources of the National Energy Research Scientific Computing Center (NERSC), a U.S. Department of Energy Office of Science User Facility located at Lawrence Berkeley National Laboratory, operated under Contract No. DE-AC02-05CH11231.

\bibliographystyle{IEEEtran}
\bibliography{references}

\end{document}